\documentclass[12pt,preprint]{aastex}

\def\omega0{\Omega_{\rm m,0}}
\def\lambda0{\Omega_{\Lambda,0}}

\def\LCDM{\Lambda{\rm CDM}}
\def\beq{\begin{equation}}
\def\eeq{\end{equation}}

\begin{document}

\title{Limits on the evolution of galaxies from the statistics of 
gravitational lenses}

\author{
Kyu-Hyun Chae\footnote{Sejong University, Department of Astronomy and Space 
Sciences, 98 Gunja-dong, Gwangjin-Gu, Seoul 143-747, Republic of Korea;
chae@arcsec.sejong.ac.kr} {}
and
Shude Mao\footnote{University of Manchester, Jodrell Bank Observatory, 
Macclesfield, Cheshire SK11 9DL, UK; smao@jb.man.ac.uk}
}

\shorttitle{Lensing limits on galaxy evolution}
\shortauthors{Chae \& Mao}

\begin{abstract}
We use gravitational lenses from the Cosmic Lens All-Sky Survey (CLASS) 
to constrain the evolution of galaxies since redshift $z \sim 1$ in the 
current $\LCDM$ cosmology. This constraint is unique as it is based on a 
mass-selected lens sample of galaxies. 
Our method of statistical analysis is the same as in Chae (2003).
We parametrise the early-type number density evolution in the form of 
$(1+z)^{\nu_n}$ and the velocity dispersion as $(1+z)^{\nu_v}$.  
We find that  
$\nu_n=-0.11^{+0.82}_{-0.89}$ ($1\sigma$) if we assume $\nu_v =0$, 
implying that the number density of early-type galaxies is
within 50\% to 164\% of the present-day value at redshift $z=1$. Allowing
the velocity dispersion to evolve, we find that $\nu_v=-0.4^{+0.5}_{-0.4}$
($1\sigma$), indicating that the velocity dispersion must be
within $57\%$ and $107\%$ of the present-day value at $z=1$. These results
are consistent with the early formation and passive
evolution of early-type galaxies.  More stringent
limits from lensing can be obtained from future large
lens surveys and by using very high-redshift quasars ($z \ga 5$)
such as those found from the Sloan Digital Sky Survey. 
\end{abstract}


\section{INTRODUCTION}

Currently there are about 70 multiply-imaged systems due to 
galactic mass scale gravitational lenses. The statistics of
gravitational lenses depend  on three key ingredients, namely, the
cosmology, the number density of potential lenses as a function of
redshift and 
the dynamical properties of galaxies (e.g., velocity dispersions and
the surface mass densities). Gravitational lenses hence encode
information of the cosmology, the galaxy mass profiles and the evolution
history of galaxies.
At present, the lens sample is too small to constrain all the
ingredients simultaneously. Most previous studies concentrated on
constraining the cosmological
constant assuming non-evolving populations of lenses (e.g., 
Fukugita et al.\ 1992; Kochanek 1996; Helbig et al.\ 1999). 
Under this assumption,
the most recent lens statistics study of Chae et al.\ (2002) finds that
the lens statistics are best-fitted by a present-day matter density of 
$\omega0 \approx 0.3$ and a cosmological constant
of $\lambda0 \approx 0.7$.
This result is consistent with results from a variety of other
studies, including the cosmic microwave background radiation
(e.g., de~Bernardis et al.\ 2000; Spergel et al. 2003), Type~Ia supernovae at 
cosmological distances (e.g., Riess et al. 1998; Perlmutter et al. 1999), 
and the large-scale structures in the universe (e.g., Peacock et al. 2001).

In light of the convergence of the cosmological model, it becomes
important to use gravitational lenses for a different purpose:
to study the evolution of galaxies
(their number density and dynamical properties) in the
$\omega0=0.3, \lambda0=0.7$ cosmology (hereafter $\Lambda$CDM). 
Lensing limits on galaxy evolution have been
explored by Mao (1991), Mao \& Kochanek (1994), Rix et al.\ (1994), and 
Jain et al.\ (2000), all of which used optically-selected lenses.
Lensing is sensitive to the evolution of galaxy properties as the
lensing probability is $\propto n\sigma^4$, while separations
are $\propto \sigma^2$; here $n$ is the number density 
and $\sigma$ is the velocity dispersion of typical lenses.
For example, (for a fixed cosmological model) a decreasing number density
of galaxies with redshift($z$) lowers the lensing rate and the mean redshift
of lenses while a decreasing velocity dispersion with $z$ lowers the 
lensing rate, the mean lens redshift and the mean angular size of image 
separations. Therefore, through a careful analysis of the lens redshifts, 
image separations and lensing probability we can constrain the evolution of 
the number density and dynamical properties of galaxies.

Most lensing galaxies are massive early-type galaxies as they dominate the 
lensing cross-sections due to their larger central mass concentrations. 
Gravitational lenses therefore provide a unique mass-selected sample to 
study the evolution of early-type galaxies,
independent of and complementary to the traditional redshift surveys of 
galaxies (e.g., Fried et al. 2001; Im et al. 2002). This is a much
debated research area. There exist two different views on the formation
and evolution of early-type galaxies, namely a 
monolithic collapse model (Eggen, Lynden-Bell, \& Sandage 1962)
and a merger hypothesis (Toomre \& Toomre 1972). In the monolithic collapse 
model, early-type galaxies are thought to have formed rapidly at high 
redshift and then evolve passively to the present-day. The merger model is 
a natural consequence of the hierarchical structure formation theory.
Semi-analytic implementations of this model predict a continuous formation 
of ellipticals and hence a certain fraction of massive 
early-type galaxies must have formed since $z \sim 1$ 
(e.g.\ Kauffmann 1996; Baugh et al.\ 1996; Kauffmann et at.\ 1999);
the fraction depends on the assumed cosmology and other assumptions and
is typically one third or more (for more, see \S4). 
An observational way of probing galaxy evolution is using redshift survey 
of galaxies. However, no consensus has been reached either with this method. 
For example, Kauffmann, Charlot, \& White (1996) found rapid evolutions 
in the number density of ellipticals while Schade et al.\ (1999) advocated 
the opposite conclusion using similar samples (see \S4 for more details).

The purpose of this work is to use data from the Cosmic Lens All-Sky 
Survey (CLASS) to provide independent constraints on galaxy evolution.
As we were completing this work, a complementary study
has been carried out by Ofek, Rix, \& Maoz (2003); their results are compared
with our results in \S2.

\section{DATA, METHOD AND RESULTS}

We use data from the Cosmic Lens All-Sky Survey (CLASS) for our study.
The survey is described extensively in Myers et al.\ (2003) and
Browne et al.\ (2003). We refer the readers to those papers for details, 
and here we only give a brief summary. The CLASS well-defined statistical 
sample contains 8958 radio sources including 13 multiply-imaged 
sources.\footnote{The rest of $\sim 7000$ CLASS sources contain further 9 
multiply-imaged sources which, however, do not fall into our statistical
sample (Myers et al. 2003; Browne et al. 2003).} 
The advantage of the CLASS survey is that it uses very well-defined 
observational selection criteria and does not suffer from the effect of
dust extinction in lenses. It is also the largest completed survey for
gravitational lenses, so it is the best sample for our purposes.

We assume the galaxy population is described by a Schechter
luminosity function (LF),
\beq
n(L)~d\left({L \over L_\star}\right) = n_\star \left({L\over
L_\star}\right)^{\alpha} \exp(-L/L_\star) d\left({L \over L_\star}\right).
\label{eq:schechter}
\eeq
Lensing galaxies are modelled as singular isothermal
ellipsoids, which are described by two parameters, the velocity 
dispersion and the axial ratio (or equivalently, the ellipticity).
The galaxy luminosity is related to the velocity dispersion via
\beq \label{eq:sigma}
{L \over L_\star} = \left({\sigma \over \sigma_\star}\right)^\gamma.
\eeq
We divide galaxies into two populations, namely the early-type 
(ellipticals and S0's) population and the late-type population and 
assume that each population is described by its own LF. 
The parameter values we take are identical to those in
Chae et al. (2002) (see also Chae 2003 for the details of the analysis 
and further results); we refer the readers to those two papers. 
In particular, we adopt the type-specific LFs based on
the galaxy classifications from the Second Southern Sky Redshift Survey
(SSRS2: Marzke et al.\ 1998); see Chae (2003) for the details. 
Our adaptation of maximum likelihood analyses (Kochanek 1993) is identical
to that of Chae et al.\ (2002) and Chae (2003). Under the $\LCDM$ cosmology
we test two simple models of the evolution of early-type galaxies.
The evolution of late-type galaxies is not considered, as
we cannot obtain any useful limits due to the small number of 
late-type lens galaxies in the current sample (i.e.\ 1 or 2).

In the first model we adopt, we assume that the shape of the 
luminosity function ($\alpha$) is a constant given by the SSRS2 (Chae 2003)
and the non-evolving (characteristic) velocity dispersion $\sigma_\star$ 
(defined in eq. \ref{eq:sigma}) is a constant to be determined from the data,
but the (characteristic) number density of galaxies  
(defined in eq.\ \ref{eq:schechter}) evolves as a function of redshift; 
we choose a power-law evolutionary shape of $1+z$,
\beq \label{eq:nz}
n_\star(z) = n_{\star,0} (1+z)^{\nu_n},
\eeq
where $n_{\star,0}$ is the present-day value.
The no-evolution model corresponds to $\nu_n=0$.
{Fig.}~1 shows confidence limits on the parameter $\nu_n$ and $\sigma_\star$.
The $\chi^2$ in {Fig.}~1 refers to $-2 \ln {\mathcal L}$ where 
${\mathcal L}$ is the likelihood function (Chae et al.\ 2002; Chae 2003).
As one can see, the lens statistics are consistent with a no-evolution model
in a $\LCDM$ cosmology at $1\sigma$ level.
We find $\nu_n = -0.11^{+0.82}_{-0.89}$; namely,
the number density of galaxies at redshift $z=1$
cannot be smaller by a factor of two or larger by $64\%$ than the 
present day number density. From {Fig.}~1 the non-evolving characteristic 
velocity dispersion is $\sigma_\star = 199^{+19}_{-16}$ 
km~s$^{-1}$($1\sigma$). This value is in good agreement with the values
from recent analyses of lensing statistics assuming no evolution of 
galaxies (Chae et al.\ 2002; Chae 2003; Davis, Huterer, \& Krauss 2003).

In the second model, we allow  the velocity dispersion $\sigma_\star$ 
(as well as the number density) to vary as a function of redshift:
\beq \label{eq:nv}
\sigma_\star(z) = 
\sigma_{\star,0} (1+z)^{\nu_v},
\eeq
where $\sigma_{\star,0}$ is the present-day characteristic velocity dispersion.
The no-evolution model corresponds to $\nu_v=0$. 
{Fig.}~2 shows the limits in the parameter space of $\nu_n$, $\nu_v$ and 
$\sigma_{\star,0}$. {Fig.}~2(a), (b) and (c) are the three projected parameter 
planes. {Fig.}~2(d) shows the limits in the plane of $\nu_v$ and $\sigma_{\star,0}$ 
based only on the image separations and the available lens redshifts 
(6 of them) of the nine single-galaxy induced multiply-imaged systems
(see Section~3.1 of Chae 2003), namely without using the lensing rate 
[see below for a discussion of the {Fig.}~2(d)].
The contours shown on each plane represent 68\%, 90\%, 95\% and 99\% 
confidence levels for one parameter.

The contours in {Fig.}~2(a) are elongated parallel to a line
$\nu_n+4\nu_v={\rm constant}$. This is because along the
$\nu_n+4\nu_v={\rm constant}$ line the optical depth ($\propto n\sigma^4$)
is a constant. In other words, there is a degeneracy in determining the
evolutions of the velocity dispersion and the number density from the optical
depth alone. Notice, however, that the degeneracy is in part broken by
the observed image separations as a function of redshift.
The $1\sigma$ limits on the two evolutionary indices are:
$\nu_n=0.7^{+1.3}_{-1.4}$ and $\nu_v=-0.4^{+0.5}_{-0.4}$.
The limit on $\nu_v$ is particularly interesting: lensing
statistics demand that the velocity dispersion for an $L_\star$ galaxy 
at $z=1$ must be between $57\%$ and $107\%$ of the present-day value. 
This implies that dynamically, the population of lensing galaxies 
cannot be much different from the present-day population. {From} Fig.\ 2(b) or
(c) the characteristic velocity dispersion of 
the present-day early-type population is 
$\sigma_{\star,0} = 223^{+38}_{-36}$~km~s$^{-1}$ ($1\sigma$).
This value has a relatively large uncertainty and is consistent with 
the value for the non-evolving case shown in {Fig.}~1.
However, it is of interest to note that the best-fit value of  
$\sigma_{\star,0}$ is somewhat larger than the non-evolving value. This is 
then consistent with the best-fit value of $\nu_v$ being negative. 

Recently, Ofek et al.\ (2003) have used the redshifts of the 
lensing galaxies in moderate-size source-redshift($z_s$)--limited 
samples to constrain the galaxy mass evolution. 
Our work is different as we use the well-defined uniform
CLASS sample and we include all the lensing information (lensing rate, image
separations and lens redshifts). Notice that our sample (13 lenses in total)
includes 6 systems with both lens and source redshifts measured while
the Ofek et al.\ (2003) samples have up to 17. They also
conclude that there is little evidence for rapid evolution of
early-type galaxies. They parameterize the evolutions in
different forms from ours. But equivalently, they find that
at 95\% confidence level, $\sigma_\star$ at $z=1$ should be at least
63\% of the present value; this is similar to our $1\sigma$ limit.
Despite the small number of the measured redshifts in our 
sample our limits on the evolution of $\sigma_\star$ are relatively 
strong because of the additional constraint of the lensing rate. 
This can be seen from the comparison of {Fig.}~2(c) and (d). 
Without the lensing rate, we have $\nu_v=0.2^{+0.6}_{-1.0}$ [{Fig.}~2(d)], 
which is significantly broader. 
The Ofek et al.\ (2003) limit on the number density evolution is given by
$d\log_{10} n_\star(z)/dz = +0.7^{+1.4}_{-1.2}$, translating into
a $1\sigma$ lower limit of the number density of lenses at $z=1$ 
of 30\% of the present value; our limit (57\%) is significantly
stronger because of the strong effects of the lensing rate.
Our results are consistent with an 
early-formation/passive-evolution picture (e.g.\ $z_{\rm formation} \ga 2$) 
of early-type galaxies, as also inferred from studies of the fundamental plane
of lensing galaxies (see Kochanek et al.\ 2000; Rusin et al.\ 2003).

\section{DISCUSSIONS}

We have used the statistical properties of the CLASS strong lens sample
(i.e., the rate of multiple-imaging and the image separations as a function
of redshift) to constrain the evolution of galaxies. The lens sample is
unique as it is mass-selected and hence the constraints obtained from it
are independent of those from redshift surveys. The method we use
is based on Chae et al.\ (2002) and Chae (2003) and has some
uncertainties, such as the adopted luminosity function of early-type
galaxies and the redshift distribution of the source
population in the CLASS survey (see Chae 2003 for details). However, 
these uncertainties are smaller than those arising from the
moderate-size CLASS sample of lenses.

We find that the (comoving) number density of lensing galaxies
at redshift $z=1$ must be within 50\% to 164\% of the
present-day number density; their characteristic velocity
dispersion also must be within $57\%$ and $107\%$ of the
present value. The lensing statistics are therefore consistent
with a slow evolution of galaxies in both their
number density and their dynamical properties.
These results are inconsistent with very fast evolution of early-type 
galaxies, where the number of early-types at $z = 1$ is 
only 20\%-40\% or less of the present-day values  
(e.g., Lin et al. 1999; Fried et al. 2001; Wolf et al.\ 2003). 
Our results are, however, consistent with several other studies, 
in particular the study based on the Hubble Space Telescope observations 
of the Groth Strip (Im et al. 2002) where there find a number
density evolution of $n(z) \propto (1+z)^{-0.86\pm 0.68}$ (see their
Table 6)  in the same underlying cosmology. In the Standard
Cold Dark Matter model with $\omega0=1$, the number density of 
bright ellipticals at $z=1$ is predicted to be a factor of 2-3 smaller 
than the local value (Kauffmann 1996; Baugh, Cole \& Frenk 1996),
inconsistent with our results. However, the evolution of galaxies
is expected to be slower in the $\LCDM$ cosmology. Indeed, 
our results are consistent with the modest evolution out to redshift
$\sim 1$ predicted by Kauffmann et al.\ (1999; see their {Fig.}~9)
for the same cosmology. Quantitatively, for ellipticals with a
stellar mass $\ga 10^{10} M_\odot$, the space density at
$z=1$ is about 30\% lower than the present day value in the fiducial
model of Cole et al.\ (2000). This result is robust to changes in the
amount of star formation occurring in bursts at high redshift (C. Baugh,
private communication; Baugh et al.\ 1996). Our results are consistent
with these predictions.

The effects of galaxy evolution on the lens statistics will be
more dramatic for very high-redshift quasars ($z \ga 5$, Fan et
al. 2003). For a source at $z=5$, half of the multiple-imaging 
cross-section (with separations between 0.3 to 5 arcseconds)
is contributed by galaxies with redshifts greater than 1.2
if there was no evolution of galaxies. Hence the lensing 
probability will be reduced significantly if the comoving 
number density of early-type galaxies is much smaller at redshifts
$z \ga 1$ compared with the local number density. A large sample
of very high-redshift quasars is therefore an independent and 
effective way of probing galaxy evolution.

The lensing constraints on the galaxy evolution are already quite
competitive compared with those from other methods. However, they still 
suffer from the small number of lenses ($13$) in our CLASS statistical
sample. With planned upgrades in major radio instruments (such
as eVLA and e-Merlin), it is possible to obtain a radio lens sample 
that is an oder of magnitude larger. When such a sample
becomes available, the lensing constraint will become more stringent
and more physical evolution (rather than our toy) models can be tested.

\acknowledgments
We thank I. Browne and C. Baugh for discussions and the anonymous referee
for constructive comments. KHC acknowledges support from
the Astrophysical Research Center for the Structure and Evolution of the 
Cosmos (ARCSEC) which was established under the KOSEF SRC program.
KHC also acknowledges the hospitality provided by the Jodrell Bank 
Observatory during the summer of 2003.

{}

\begin{figure}
\epsscale{0.8}
\plotone{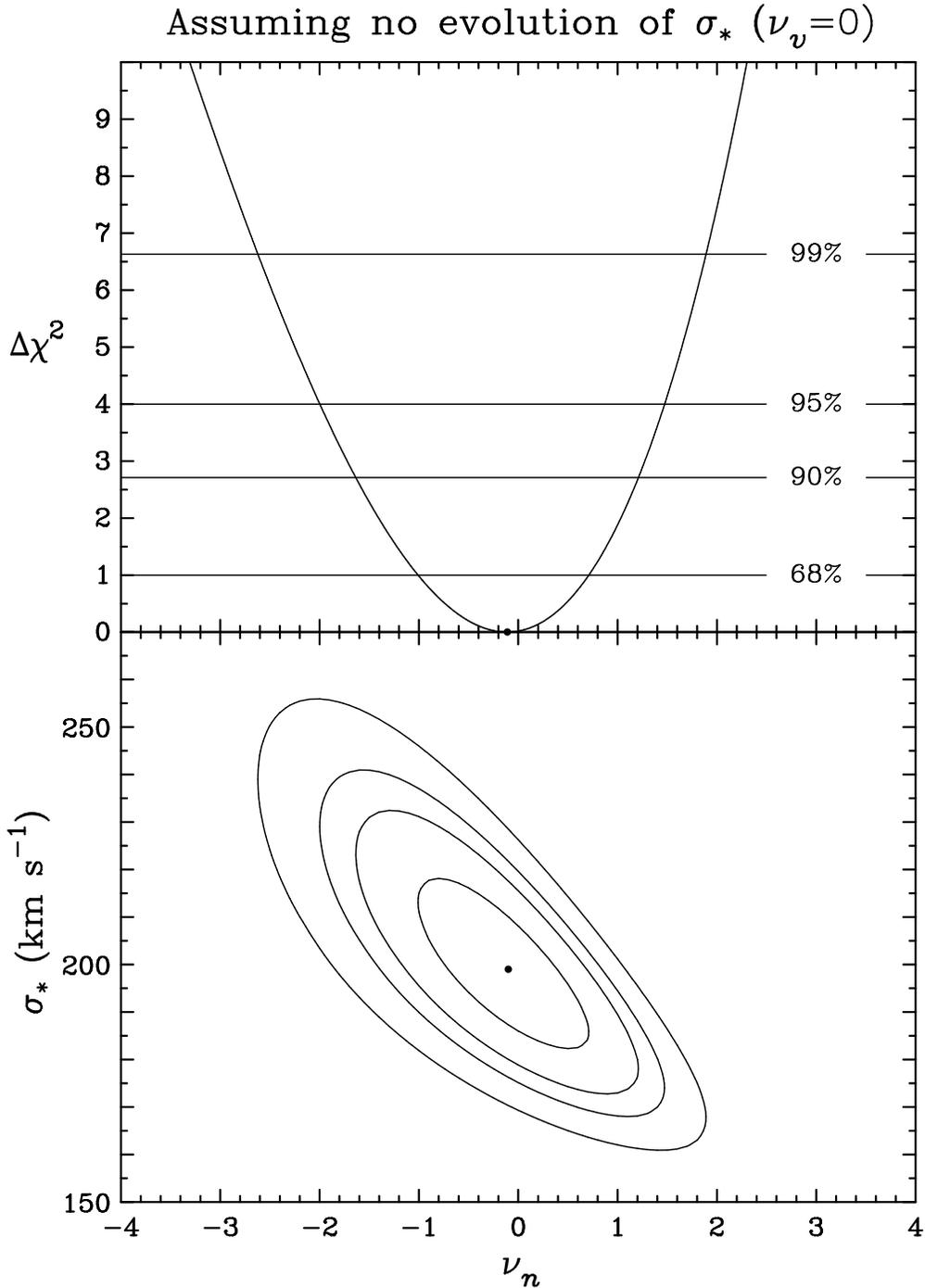}
\caption{
Confidence limits on the number-density evolutionary index 
for early-type galaxies, $\nu_n$ (see eq. \ref{eq:nz}), assuming no evolution
of the velocity dispersion.  The lower panel shows the likelihood
contours in the $\nu_n$ and $\sigma_\star$ plane, with
the solid dot indicating the peak of the likelihood function.
The upper panel shows $\Delta\chi^2$ as a function of $\nu_n$ where
we have marginalized $\sigma_\star$. Here $\chi^2$ is defined as
$-2\ln {\mathcal L}$ where ${\mathcal L}$ is the likelihood function.
}
\label{fig:fig1}
\end{figure}

\begin{figure}
\epsscale{0.8}
\plotone{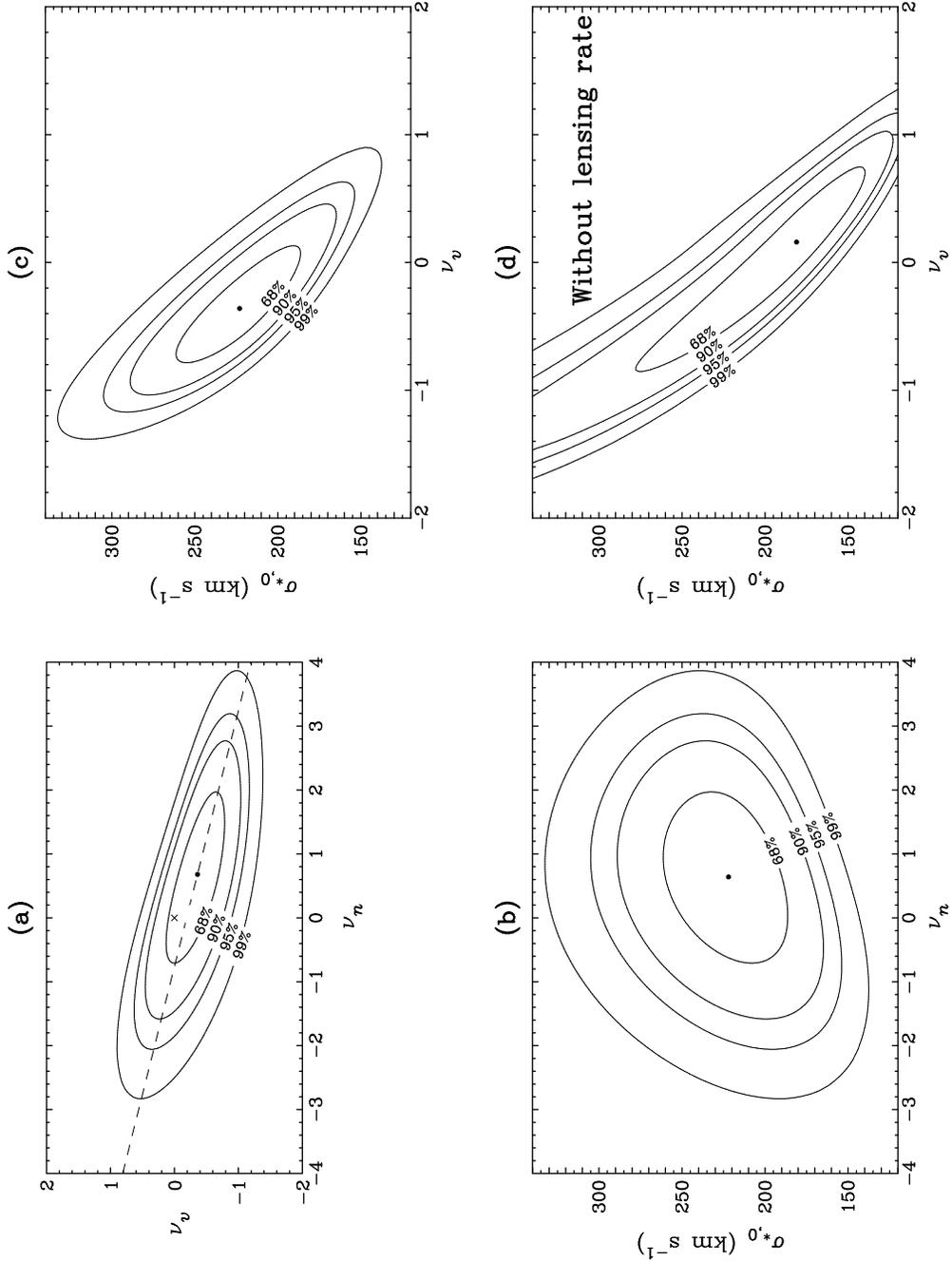}
\caption{
Confidence limits on the number density evolution index,
$\nu_n$, the velocity dispersion evolution index, $\nu_v$ (defined 
in eqs. \ref{eq:nz} and \ref{eq:nv}), and the present-day velocity 
dispersion $\sigma_{\star,0}$.
Panels (a), (b) and (c) show three different projections in
the $\nu_n$-$\nu_v$, $\nu_n$-$\sigma_{\star,0}$ and
$\nu_v$-$\sigma_{\star,0}$ planes, respectively; the third
remaining parameter has been marginalized. In each panel,
the solid dot indicates the peak of the likelihood function.
In Fig.\ 2(a), the dashed line shows the line where the optical depth is
kept as a constant ($\nu_n+4\nu_v={\rm constant}$), while 
the origin (marked by a cross) corresponds to the no-evolution case.
Fig.\ 2(d) shows the likelihood contours on $\nu_v$ and $\sigma_{\star,0}$ 
where we do not incorporate the lensing rate information in our
maximum likelihood calculation.
}
\label{fig:fig2}
\end{figure}

\end{document}